%% file: main.tex
\documentclass[conference,comsoc]{IEEEtran}
\usepackage{amsmath}

\usepackage{graphicx}   
\usepackage{comment}
\usepackage{amssymb}  
\newcommand{\xmark}{\ding{55}}
\usepackage{wasysym}

\usepackage{mathtools}
\usepackage{aligned-overset}
\usepackage{caption}
\usepackage{pifont}
\usepackage{textcomp}
\usepackage{algpseudocode}

\usepackage{cite}

\allowdisplaybreaks
\usepackage{lipsum}

\usepackage{array}
\usepackage{booktabs}  

\usepackage{enumitem,kantlipsum}
\usepackage{multirow,hhline}
\usepackage{colortbl}
\usepackage{listings}
\usepackage{makecell}
\usepackage{hyperref}
\usepackage{xspace}
\hypersetup{colorlinks=true,
            linkcolor=black,
            anchorcolor=black,
            citecolor=black}
\usepackage{url}

\newcommand{\hz}[1]{\textbf{\textcolor{red}{[#1]}}}

\usepackage{cleveref}

\usepackage[linesnumbered,ruled,noline]{algorithm2e}

\usepackage{tikz}

\SetCommentSty{mycommfont}
\SetKwInput{KwInput}{Input}                
\SetKwInput{KwOutput}{Output}              
\SetKwBlock{Loop}{Loop}{}
\SetKwBlock{Func}{Function}{}
\SetKwBlock{If}{If}{}

\ifCLASSOPTIONcompsoc
  \usepackage[nocompress]{cite}
\else
  \usepackage{cite}
\fi

\begin{document}

\newcommand{\pname}{{\texttt{RadEar}}\xspace} 



\title{\pname: A Self-Supervised RF Backscatter System for Voice Eavesdropping and Separation\vspace{-0.15in}
}

\author{
Qijun Wang$^{\ast}$,
Peihao Yan$^{\ast}$,
Chunqi Qian$^{\dagger}$,
and Huacheng Zeng$^{\ast}$\\
$^{\ast}$Department of Computer Science and Engineering, Michigan State University\\
$^{\dagger}$Department of Radiology, Michigan State University
\vspace{-0.25in}
}



\maketitle

\begin{abstract}
\input{abstract}
\end{abstract}

\IEEEpeerreviewmaketitle

\section{Introduction}
\input{1_introduction}


\section{Threat Model and System Overview}
\input{2_threat_model}

\section{RF Backscatter Tag Design}
\input{3_RF_backscatter_tag_design_for_multi-speaker_voice_recovery}

\section{RF Reader Design: Audio Recovery and Separation}

\input{4_audio_denoise_and_separation}

\section{Experimental Evaluation}
\input{5_evaluation}

\section{Related Work}
\input{6_related_work}


\section{Conclusion}
\input{7_conclusion}

\section*{Acknowledgment}
This work was supported in part by NSF Grant ECCS-2225337 (Q.~Wang, P.~Yan, and H.~Zeng) and in part by NSF Grant ECCS-2144138 and NIH Grant RF1NS128611 (C.~Qian).

\bibliographystyle{IEEEtran}
\bibliography{references} 

\end{document}

%% file: abstract.tex
Eavesdropping on voice conversations presents a growing threat to personal privacy and information security. In this paper, we present \pname, a novel RF backscatter-based system 
designed to enable covert voice eavesdropping through walls. 
\pname consists of two key components: (i) a batteryless RF backscatter tag covertly deployed inside the target space, and (ii) an RF reader located outside the room that performs signal demodulation, voice separation, and denoising. The tag features a compact, dual-resonator design that achieves energy-efficient frequency modulation for \textit{continuous} voice eavesdropping while mitigating self-interference by separating excitation and reflection frequencies. 
To overcome the challenges of weak signal reception and overlapping speech, the RF reader employs self-supervised learning models for voice separation and denoising, trained using a remix-based objective without requiring ground-truth labels. We fabricate and evaluate \pname in real-world scenarios, demonstrating its ability to recover and separate human speech with high fidelity under practical constraints. 

\begin{IEEEkeywords}
    Audio security, eavesdropping, RF backscatter communication, self-supervised learning, information privacy
\end{IEEEkeywords}

%% file: 1_introduction.tex
Eavesdropping poses a critical threat to our society as it undermines the fundamental right to privacy and compromises the confidentiality of personal and business conversations. In an era where voice-enabled technologies are deeply embedded in everyday life, unauthorized access to spoken conversations can lead to identity theft, intellectual property loss, corporate espionage, and even national security breaches. Unlike traditional data breaches, voice eavesdropping is particularly insidious because it can occur passively and covertly, without leaving digital traces. As communication technologies evolve, so do the capabilities of adversaries to exploit wireless technologies and physical environments to capture sensitive information from a distance. This escalating threat demands a deeper understanding of possible eavesdropping technologies and their attack surfaces, enabling researchers and policymakers to develop countermeasures that safeguard voice privacy.

Since voice-based acoustic signals are easily blocked by soundproof materials, direct acoustic eavesdropping from outside a private space is often infeasible. As a result, Radio Frequency (RF)-based communication and sensing technologies have been explored in increasingly sophisticated forms for eavesdropping applications. Existing efforts include mmWave-based voice detection \cite{wang2022wavesdropper,zhao2023radio2text,zhang2022ambiear,ozturk2022radiomic,wang2022mmeve,shi2023privacy}, RFID-based voice detection \cite{chen2024rfspy,yang2024rf, wang2025batteryless}, and UWB-based voice detection \cite{wang2024vibspeech}. 
While these systems demonstrate the potential of RF signals to recover speech, their practical applications are constrained to controlled environments and lack robustness in real-world scenarios.
For example, mmWave signals poorly penetrate walls due to high-frequency attenuation \cite{zhang2025radsee, zhang2025radeye}, RFID-based methods require physical proximity to the speaker, and UWB approaches face challenges with multipath interference and synchronization.


\begin{figure}
\centering
 \includegraphics[width=0.985\linewidth]{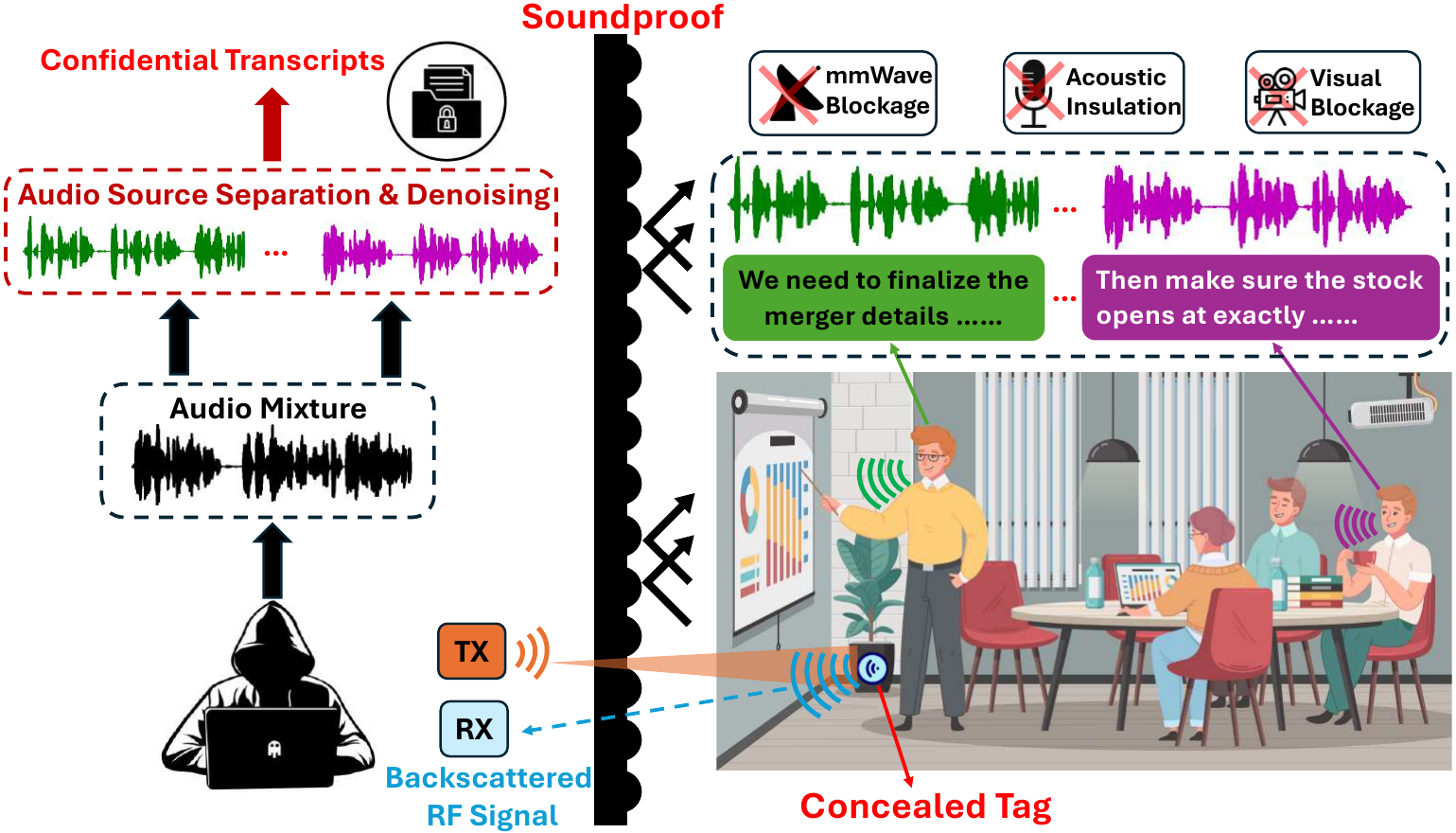}
 \vspace{-0.025in}
\caption{Threat model and system configuration.}
\vspace{-0.15in}
\label{fig:scenario}
\end{figure}

In this paper, we consider a threat model illustrated in Fig.~\ref{fig:scenario}, where a confidential conversation involving one or multiple people takes place in a private room protected by soundproof walls.
An adversary, located outside the target space, aims to eavesdrop on the conversation without physical access.
Moreover, the adversary seeks to separate individual speakers' voices to enable automatic speech transcription, paving the way for advanced semantic analysis such as speaker identification, intent recognition, or topic extraction.
To execute this attack, we present an RF backscatter system called \underline{Rad}io \underline{Ear}, or \pname for short.
\pname comprises two main components:
(i) a passive RF backscatter tag covertly placed within the target space, and
(ii) an RF reader positioned outside the room to capture and decode the RF signals modulated by the ambient acoustic vibrations of human speech.

To support a reliable eavesdropping task, the design of the RF backscatter tag must address several key challenges.
\textit{First}, the tag must be batteryless and operate solely on harvested energy for computation and communication, enabling permanent deployment within the target space. Moreover, the tag must be small (e.g., less than one inch) and unobtrusive to allow for easy concealment (e.g., covertly attached under a table or chair).
\textit{Second}, the tag must support \textit{continuous} voice streaming. While prior batteryless RF backscatter tags have demonstrated voice transmission capabilities (e.g., WISP~\cite{menon2022wireless}, Battery-Free Phone~\cite{talla2017battery}, MARS~\cite{arora2021mars}), they are incapable of supporting \textit{continuous} voice streaming, as they rely on long-time energy harvesting for short-time operation.
\textit{Third}, the tag must reliably modulate the RF signal to maximize the eavesdropping range. A notorious issue in many RF backscatter systems (e.g., RFID) is self-interference: when the tag modulates the voice signal on the same frequency as the excitation signal, the RF reader experiences strong self-interference, significantly degrading its detection range.

To address these challenges, we introduce a novel design for the RF backscatter tag. 
Our proposed tag comprises four elements: 
a piezoelectric sensor, 
a voltage-sensing resonator (VSR), 
a parametric resonator (PR), 
and 
a dipole antenna.
The piezoelectric sensor transduces voice-induced pressure fluctuations into voltage signals, which modulate the VSR’s resonance frequency.
The VSR is magnetically coupled to the PR, which serves as an energy pump and introduces spectral separation via voltage-tunable dual-mode resonance. The modulated signal is radiated through a dipole antenna toward the RF reader, boosting the eavesdropping range.


This new design offers four key advantages over existing backscatter tags.
First, it achieves significant spectral separation between excitation and reflection, fundamentally mitigating the issue of \textit{self-interference} at the RF reader.
Second, voice signals are modulated via analog frequency shifts, enabling \textit{continuous} voice streaming without digitization.
Third, the tag supports frequency modulation (FM), establishing a linear relationship between shifts in the tag’s resonance frequency and the amplitude of the voice signal. This greatly simplifies voice recovery at the RF reader.
Fourth, the entire design is compact enough to fit on a one-inch PCB, allowing for easy concealment within the target space.
Additionally, the tag operates at a low frequency (e.g., 915 MHz), which offers strong wall penetration and favorable propagation characteristics for through-wall eavesdropping.

To increase the eavesdropping range and enable voice separation, the RF reader plays a critical role but faces two key challenges.
First, the received signal is extremely weak due to passive modulation and wall attenuation, complicating voice recovery.
Second, voice separation is inherently challenging because there are no ground-truth labels for individual speakers' voice data. The low signal-to-noise ratio (SNR) further exacerbates this challenge, making it difficult to isolate and recover clean voice signals.

To address these issues, we propose a learning-based RF reader design that incorporates two innovative components: a voice \textit{separation} model and a voice \textit{denoising} model. Both models are trained in a self-supervised manner to improve their performance without relying on labeled data.
Our design leverages the statistical regularities present in human voice mixtures, which allow models to learn meaningful patterns without explicit supervision. Inspired by remixing-based approaches such as those in \cite{wisdom2020unsupervised,tzinis2022remixit}, we train our models to decompose voice mixtures into multiple components and use a reconstruction loss to guide training. Specifically, the separated components are recombined to reconstruct the original mixture, and the loss measures the fidelity of this reconstruction.
This remixing-based training provides an implicit form of supervision, encouraging the models to learn useful representations of the underlying sources, thereby improving both voice separation and noise suppression in weak RF signals. 

\input{tables/comparison1}

We have fabricated the RF backscatter tag and built the RF reader, and evaluated the performance of \pname in real-world environments. Extensive experiments demonstrate that \pname provides a satisfactory voice recovery and separation under various realistic conditions. 
Table~\ref{tab:comparison} summarizes the key achievements of \pname against existing approaches.

This work advances the state-of-the-art as follows.

\begin{itemize}[leftmargin=0.2in]
\item
It introduces a novel design for an RF backscatter tag capable of supporting \textit{continuous} voice streaming via \textit{frequency modulation}.

\item
It presents a new RF reader design that incorporates \textit{self-supervised} modules for voice separation and denoising.

\item
Extensive experiments validate the practicality of the proposed eavesdropping system and demonstrate its superior performance.
\end{itemize}







%% file: tables/comparison1.tex
\begin{table}[t]
\centering
\caption{
Comparison of Eavesdropping Attack Methodologies. 
“GT Req.” indicates whether ground truth data is required for training. 
Symbol legend: \Circle = Low, \LEFTcircle = Medium, \CIRCLE = High.
}\label{tab:comparison}\vspace{-0.05in}


\resizebox{0.48\textwidth}{!}{%
\begin{tabular}{c|c|c|c|c|c|c|c|c}
\hline
\multirow{2}{*}{Previous Work}   & \multirow{2}{*}{Hardware} & \multirow{2}{*}{Max. Dist.}  & \multirow{2}{*}{Thru-Wall} & \multirow{2}{*}{GT Req.} & \multirow{2}{*}{Preset Input?}  & \multirow{2}{*}{Mobility} & \multicolumn{2}{c}{Generalizability}   \\ \cline{8-9}
 &  &  &  &   &   &  & User & Env.  \\ \hline
 
WaveEar \cite{xu2019waveear}   & mmWave Radar & 2 m   & \textbf{\xmark} & \checkmark & \textbf{\xmark}   & \Circle & \LEFTcircle & \LEFTcircle  \\ \hline

AccelEve \cite{ba2020learning}   & Accelerometer & 0 m   & \textbf{\xmark} & \checkmark & \textbf{\checkmark}   & - & \LEFTcircle & \LEFTcircle   \\ \hline

UWHear \cite{wang2020uwhear}   &  IR-UWB Radar & 8 m   & \textbf{\checkmark} & \xmark & \textbf{\xmark}   & - & \LEFTcircle & \LEFTcircle   \\ \hline

Tag-Bug \cite{wang2021thru}   &  RFID Tag & 2 m   & \textbf{\checkmark} & \checkmark & \textbf{\checkmark}   & - & \LEFTcircle & \LEFTcircle   \\ \hline

AccEar \cite{hu2022accear}   &  Accelerometer & 0 m   & \textbf{\xmark} & \checkmark & \textbf{\xmark}   & \LEFTcircle & \LEFTcircle & \LEFTcircle   \\ \hline

MILLIEAR \cite{hu2022milliear}   &  mmWave Radar & 5 m   & \textbf{\checkmark} & \checkmark & \textbf{\xmark}   & - & \LEFTcircle & \LEFTcircle   \\ \hline

mmSpy \cite{basak2022mmspy}   &  mmWave Radar & 1.8 m   & \textbf{\xmark} & \checkmark & \textbf{\checkmark}   & \Circle & \LEFTcircle & \LEFTcircle  \\ \hline

mmPhone \cite{wang2022mmphone}   &  mmWave Radar & 5 m   & \textbf{\checkmark} & \xmark & \textbf{\checkmark}  &  - & \LEFTcircle & \LEFTcircle  \\ \hline

Wavesdropper \cite{wang2022wavesdropper}  &  mmWave Radar & 5 m   & \textbf{\checkmark} & \checkmark & \textbf{\checkmark}   & \LEFTcircle  & \Circle & \LEFTcircle   \\ \hline

mmEve \cite{wang2022mmeve}   &  mmWave Radar & 8 m   & \textbf{\xmark} & \checkmark & \textbf{\xmark}   & \LEFTcircle & \LEFTcircle & \LEFTcircle   \\ \hline

RADIOMIC \cite{ozturk2022radiomic}   &  mmWave Radar & 4 m   & \textbf{\checkmark} & \xmark & \textbf{\xmark}  & \LEFTcircle & \LEFTcircle & \LEFTcircle  \\ \hline

AmbiEar \cite{zhang2022ambiear}   &  mmWave Radar & 2.5 m   & \textbf{\xmark} & \checkmark & \textbf{\xmark}   & \LEFTcircle & \LEFTcircle & \LEFTcircle   \\ \hline

Radio2Text \cite{zhao2023radio2text}   &  mmWave Radar & 1.5 m   & \textbf{\checkmark} & \checkmark & \textbf{\xmark}   & - & \LEFTcircle & \LEFTcircle   \\ \hline

Shi \textit{et al}. \cite{shi2023privacy}   &  mmWave Radar & 11 m   & \textbf{\checkmark} & \checkmark & \textbf{\checkmark}  & \LEFTcircle & \LEFTcircle & \LEFTcircle   \\ \hline

mmEav \cite{feng2023mmeavesdropper}  &  mmWave Radar & 3 m   & \textbf{\xmark} & \checkmark & \textbf{\checkmark}   & - & \LEFTcircle & \Circle   \\ \hline




RFSpy \cite{chen2024rfspy}   &  RFID Tag & 3.5 m   & \textbf{\checkmark} & \checkmark & \textbf{\xmark}   & \LEFTcircle & \LEFTcircle & \LEFTcircle   \\ \hline

VibSpeech \cite{wang2024vibspeech}   &  mmWave Radar & 5 m   & \textbf{\checkmark} & \xmark & \textbf{\xmark}   & \Circle & \LEFTcircle & \LEFTcircle  \\ \hline

mmEar \cite{xu2024mmear}   &  mmWave Radar & 2 m   & \textbf{\xmark} & \checkmark & \textbf{\xmark}   & \Circle  & \LEFTcircle & \LEFTcircle  \\ \hline


RF-Parrot \cite{yang2024rf}   &  RF tag & 1 m   & \textbf{\checkmark} & \checkmark & \textbf{\checkmark}  & - & \LEFTcircle & \LEFTcircle   \\ \hline





Onishi \textit{et al}. \cite{onishi2025sound}  &  RF Transceiver  & 2 m  & \textbf{\checkmark} & \xmark & \textbf{\xmark}  & - & \LEFTcircle & \LEFTcircle  \\ \hline


\textbf{This work}   & \textbf{RF transceiver} & \textbf{8 m}  & \textbf{\checkmark} & \xmark & \textbf{\xmark}  & \CIRCLE & \CIRCLE & \CIRCLE  \\

 \hline
\end{tabular}
}
\vspace{-0.1in}
\end{table}

%% file: 2_threat_model.tex
\subsection{Threat Model}
We consider a threat model as illustrated in Fig.~\ref{fig:scenario}, where a confidential conversation involving one or multiple individuals takes place inside a private room protected by soundproof walls.  
An adversary, positioned outside the target area, aims to \textit{continuously} eavesdrop on the conversation without any physical access to the room.  
Furthermore, the adversary seeks to separate individual speakers’ voices to enable automatic speech transcription, thereby facilitating advanced semantic analysis such as speaker identification, intent recognition, and topic extraction.

To execute this attack, the adversary covertly deploys an RF backscatter tag inside the target room.  
This tag is battery-less and operates solely by harvesting energy from an external RF reader located outside the target area. 
This tag is required to operate continuously for voice streaming with a full duty cycle. 
Additionally, the tag must be small and unobtrusive, allowing it to be discreetly placed within the environment.

The adversary has no access to clean ground-truth audio data from individual speakers, no control over their positions, and no ability to physically tamper with or alter the environment beyond the initial placement of the tag.  
However, the adversary can access a location outside the target space (e.g., behind a wall) to deploy an RF reader.  
This reader is responsible for RF signal transmission, signal processing, and learning-based computation, with the goal of recovering and separating the voice streams of individual speakers.

\subsection{Proposed Eavesdropping System}

To execute this attack, we present \pname, an RF backscatter-based eavesdropping system consisting of two main components:
(i) a passive RF backscatter tag covertly placed within the target space, and
(ii) an RF reader positioned outside the room to perform signal demodulation, recovery, and voice separation.
The tag features a novel architecture with two coupled resonators (VSR and PR), which enable separation of the excitation and reflection signal frequencies, thereby mitigating self-interference for signal demodulation at the RF reader.
This architecture also supports energy-efficient frequency modulation for voice signals, enabling \textit{continuous} voice streaming on the tag.
The RF reader incorporates two learning-based modules for voice separation and denoising, both of which can be trained in a self-supervised manner to enhance performance in real-world, unlabeled environments.

The following sections detail the RF tag and reader.

%% file: 3_RF_backscatter_tag_design_for_multi-speaker_voice_recovery.tex
\begin{figure}
\centering
 \includegraphics[width=0.8\linewidth]{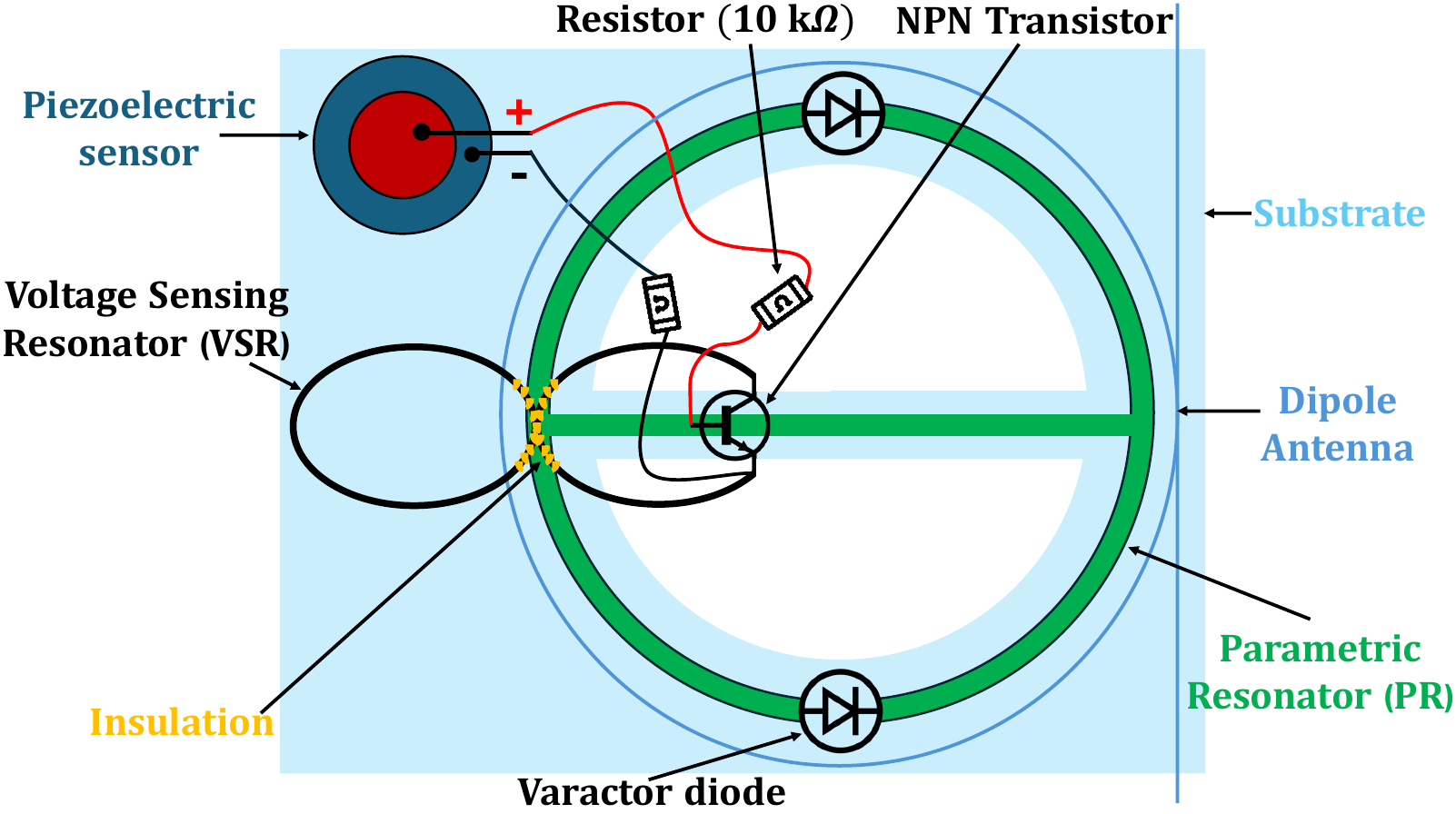}
\caption{Diagram of our RF backscatter tag.}
\vspace{-0.25in}
\label{fig:tag_arc}
\end{figure}

\subsection{Overview}

The design of a backscatter tag for continuous voice streaming faces two key challenges:
\textit{(i) Frequency Separation:}
Traditional RF backscatter tags, like RFID, reflect signals at the same frequency as their excitation source. Since the excitation signal is much stronger, it overwhelms the weak reflected signal, causing severe self-interference at the RF reader. This degrades sensitivity, limits range, and leads to high packet error rates (PER). While such degradation may be tolerable for RFID, it is unacceptable for voice streaming, which requires a reliable, continuous data link. Thus, effective separation of excitation and reflection frequencies is essential.
\textit{(ii) Efficient Voice Signal Modulation:}
RF backscatter tags harvest only minimal power, insufficient to support conventional digital voice modulation. While analog modulation is more power-efficient, existing techniques still consume too much energy for continuous streaming on batteryless tags. These limitations call for novel, low-power modulation methods designed to enable real-time voice communication over RF backscatter.

To address these two challenges, we propose an RF backscatter tag.
This RF tag is designed to operate passively, harvesting energy from an external RF reader while modulating acoustic signals onto the reflected carrier. 
As shown in Fig.~\ref{fig:tag_arc}, the tag consists of four primary components: 
a piezoelectric sensor, 
a voltage-sensing resonator (VSR), 
a parametric resonator (PR), 
and 
a dipole antenna.
The piezoelectric sensor captures pressure fluctuations induced by nearby voice activity and converts them into low-voltage electrical signals. 
These signals directly modulate the impedance characteristics of the VSR, whose resonant frequency shifts accordingly. 
The modulated VSR is coupled to the PR, which acts as an energy pump to enhance the backscatter efficiency. 
Additionally, the PR separates the excitation and backscatter/reflection signals in frequency due to the dual-mode resonance structure.
Finally, the composite signal is radiated via the dipole antenna toward the RF reader, which demodulates the acoustic content embedded in the carrier waveform.

The layout of the VSR and PR resonators, along with their coupling mechanism, is co-designed and jointly optimized to ensure high signal integrity under low-SNR and wide-angle acoustic detection conditions, enabling long-range eavesdropping capabilities.
In what follows, we describe the design and operation of the VSR and PR in detail.

\subsection{Voltage-Sensing Resonator (VSR)} 

The VSR serves as the core transduction unit that maps acoustic pressure into modulated RF backscatter. 
To support high-sensitivity operation under passive constraints, we implement a novel BJT-based resonator structure that achieves frequency modulation through piezo-induced voltage variations.

The resonator consists of an $\infty$-shaped metallic conductor with a central symmetry axis, terminated by a bipolar junction transistor (NPN-type) across its open ends. 
The base and emitter terminals of the transistor are connected to the differential output of a piezoelectric diaphragm, while the collector remains electrically floating. 
This configuration introduces two PN junctions in series, forming a voltage-variable capacitive network embedded within the resonant loop.

When an incident acoustic waveform induces pressure \( p(t) \) on the piezo element, it generates a voltage \( v_{\mathrm{pz}}(t) \propto p(t) \) across the base-emitter junction. 
This voltage modulates the depletion width \( w(t) \) of each PN junction, thereby altering its junction capacitance:

\begin{equation}
C_{\mathrm{PN}}(t) = \frac{\varepsilon A}{w(t)} \approx \frac{C_0}{\sqrt{1 + v_{\mathrm{pz}}(t)/\phi_T}},
\label{eq:cpn}
\end{equation}
where \( C_0 \) is the junction capacitance at equilibrium, \( \phi_T \) is the thermal voltage, and \( \varepsilon A \) represents the junction permittivity-area product. 
The total capacitance of the resonator becomes a function of the input sound pressure. 
Denote $f_{\mathrm{res}}(t)$ as the instantaneous resonance frequency of VSR. 
Then, based on Eqn~\eqref{eq:cpn}, we have:
\begin{equation}
f_{\mathrm{res}}(t) 
= \frac{1}{2\pi \sqrt{L C_{\mathrm{PN}}(t)}}
\approx \frac{1}{2\pi \sqrt{LC_0}}  \Big(1 + v_{\mathrm{pz}}(t)/\phi_T \Big)^{\frac{1}{4}}.
\label{eq:fres}
\end{equation}

In practice, the voice-induced piezo voltage is weak and thus we have 
$v_{\mathrm{pz}}(t)/\phi_T \ll 1$. 
Therefore, based on the Taylor series expansion, 
Eqn~\eqref{eq:fres} can be approximated by:
\begin{equation}
f_{\mathrm{res}}(t) 
\approx \frac{1}{2\pi \sqrt{LC_0}} \Big(1 + \frac{v_{\mathrm{pz}}(t)}{4\phi_T}\Big).
\label{eq:fres2}
\end{equation}

Eqn~\eqref{eq:fres2} shows the linear relationship between VSR's resonance frequency $f_{\mathrm{res}}(t)$ and the input sound pressure $v_{\mathrm{pz}}(t)$. 
This is Frequency Modulation (FM). 
It encodes the temporal variations of voice signals directly into the tag's resonant frequency, which is later extracted by the RF reader.

In our design, the bipolar transistor's dual-junction structure enables symmetric capacitance tuning, which improves linearity and reduces phase noise in VSR's frequency modulation. Additionally, the $\infty$-shaped conductor suppresses common-mode interference, spatially varying environmental noise, and multipath artifacts. Together, these features enhance the robustness and stability of the VSR's frequency modulation across diverse indoor deployment scenarios.


\subsection{Parametric Resonator (PR)} 
 
\textbf{Separation of Excitation and Reflection Frequencies:}
Traditional backscatter systems suffer from self-interference due to spectral overlap between excitation and reflection signals, making it difficult for the RF reader to recover weak modulated signals.
To address this, we propose a dual-mode PR, as illustrated in Fig.~\ref{fig:tag_arc}, which supports concurrent circular and butterfly resonance modes.
Let $f_c$ and $f_b$ denote the resonance frequencies of the circular and butterfly modes, respectively.
When the tag is excited by an external signal at frequency
$f_{\text{ex}} = f_c + f_b$,
both modes are simultaneously activated, generating signals at $f_c$ and $f_b$.
We designate $f_c$ as the carrier frequency of the reflected signal and enhance its radiation with a dipole antenna.
This separation of excitation and reflection frequencies allows the RF reader to effectively isolate the backscattered signal and suppress self-interference.

\textbf{Magnetic Coupling Between PR and VSR:}
The VSR is magnetically coupled to the PR's circular resonance mode, allowing voice-induced voltage shifts to modulate the reflection frequency via frequency modulation. The butterfly mode in the PR facilitates energy pumping and oscillation without interfering with voice encoding. This dual-mode coupling scheme enables passive voice modulation and amplification while maintaining spectral isolation.
The resonant coupling further enables efficient transfer of modulated energy to the antenna, enhancing backscatter signal strength while preserving the tag's passive nature.

To improve voice-induced frequency modulation efficiency, the PR is designed as a lumped-element LC circuit incorporating a voltage-tunable varactor diode. The resonance frequency $f_{\mathrm{c}}$ of its circular mode is tuned to closely match the nominal resonance frequency of the VSR. 
When the two resonators are placed in close proximity, they are electromagnetically coupled via mutual inductance, enabling energy exchange.

By carefully tuning the varactor in the PR, we bias its resonance slightly off the carrier frequency to maximize sensitivity to the frequency-shifted sidebands induced by voice signals. This asymmetric tuning allows the PR to function as a passive, frequency-selective amplifier, concentrating energy from the voice-modulated subcarriers into a narrow spectral band aligned with the reader's receiving frequency.

\textbf{Dipole Antenna:}
A dipole antenna is wrapped around the edge of the PR’s circular conductor to enhance inductive coupling.
Resonant amplification allows small variations in the VSR’s frequency to produce detectable amplitude and phase shifts in the backscattered signal, enabling robust acoustic signal recovery even under low-SNR indoor conditions.

\textbf{Backscattered Signal:}
In principle, the PR and the dipole antenna together function as an amplifier, enabling the VSR to more effectively radiate its modulated signal.
Coarsely, the instantaneous frequency of the tag’s backscattered signal can be approximated as:
\begin{equation}
f_c(t) 
\approx 
\gamma_1 \gamma_2 f_\mathrm{res}(t)
\approx 
 \frac{\gamma_1 \gamma_2}{2\pi \sqrt{LC_0}} \left(1 +  \frac{v_{\mathrm{pz}}(t)}{4\phi_T} \right),
\label{eq:backscatter}
\end{equation}
where $\gamma_1$ and $\gamma_2$ are the gains from the PR and dipole antenna, respectively.
Overall, the coupled VSR–PR system acts as a compact, passive frequency modulation transducer that enhances acoustic detection sensitivity, forming the physical foundation for long-range voice detection.



%% file: 4_audio_denoise_and_separation.tex
\begin{figure}
\centering
 \includegraphics[width=1\linewidth]{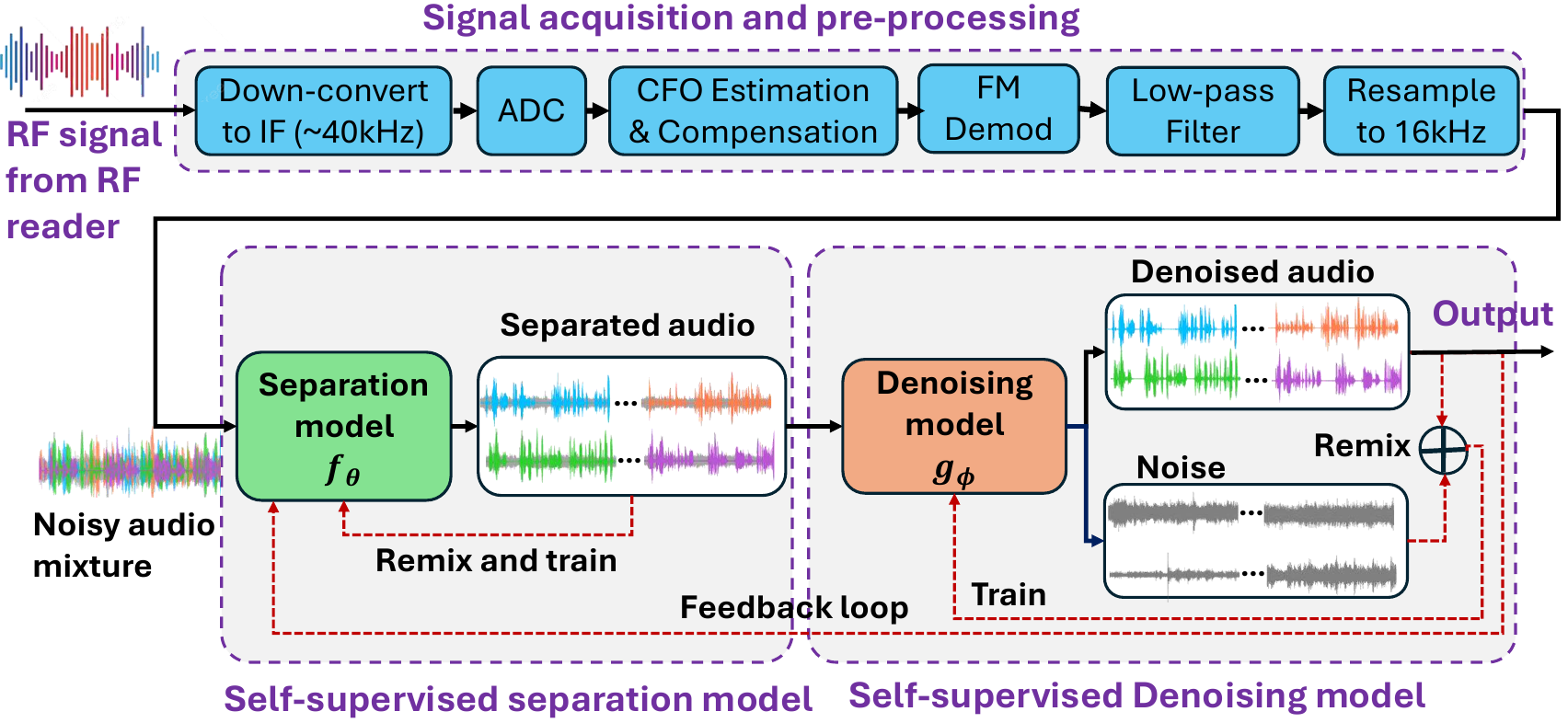}
 \vspace{-0.15in}
\caption{Diagram of our RF reader design.}
\vspace{-0.15in}
\label{fig:system_arc}
\end{figure}

The RF reader plays a critical role in demodulating, separating, and denoising the voice signal received from the backscatter tag.
Fig.~\ref{fig:system_arc} illustrates our RF reader design, which comprises three key components:
(i) signal acquisition and pre-processing,
(ii) a self-supervised model for voice separation, and
(iii) a self-supervised model for voice denoising.
In the following sections, we describe each component in detail.

\vspace{-0.05in}
\subsection{Signal Acquisition and Pre-Processing}
\vspace{-0.05in}

To detect the acoustic signal, the RF reader performs continuous-wave (CW) transmission at a fixed carrier frequency \( f_{\mathrm{ex}} \).
Meanwhile, it receives the backscattered signal from the tag at a different frequency $f_\mathrm{c}$.
The isolation of the transmitting and receiving signal frequencies avoids the notorious self-interference issues and thus significantly improves the voice detection range. 
The backscattered RF signal is first down-converted to an intermediate frequency (IF) (e.g., 40 kHz) and then sampled for digital signal processing.
The reason why we convert the radio signal to IF rather than baseband (zero-IF) is twofold. 
First, voice signals contain rich low-frequency components (typically below 100 Hz), and the RF front end often suffers from disturbances near the DC component.
Therefore, using IF will avoid interference at low frequencies. 
Second, the radio frequency suffers from CFO. 
Since the subsequent process needs to compensate the CFO, using IF will not incur an additional processing burden.

The digitalized signal samples are then processed by a sequence of blocks for CFO compensation. 
In this process, the CFO is tracked over time and compensated for individual segmented signal frames. 
The signal after CFO compensation can be expressed as:
\begin{equation}
r(t) = A(t) \cdot \exp\big(j 2\pi \cdot f_\mathrm{bb}(t) \cdot t \big) + n(t),
\end{equation}
where $A(t)$ represents time-varying signal amplitude and \(  f_\mathrm{bb}(t) \) represents the instantaneous frequency deviation introduced by the acoustic signal, and \( n(t) \) captures RF noise.
We demodulate the voice signal by differentiating the phase of $r(t)$ over time. 
The demodulated voice signal, denoted as $x(t)$, can be written as:
\begin{equation}
    x(t) = \frac{f_\mathrm{s}}{2\pi \Delta} [\angle r(t + \Delta) - \angle r(t)],
    \label{eq:recoveredsignal}
\end{equation}
where $f_\mathrm{s}$ is the sampling rate of $r(t)$ and $\Delta$ is a fixed small time step. 
We note that the demodulated voice signal $x(t)$ in Eqn~\eqref{eq:recoveredsignal} is an estimate of 
$v_\mathrm{pz}(t)$ in Eqn~\eqref{eq:cpn}. 
The demodulated signal $x(t)$ serves as the input to the downstream voice separation and denoising pipeline.








\vspace{-0.05in}
\subsection{Self-supervised Audio Separation Model}
\vspace{-0.05in}

The demodulated signal contains a mixture of speech from multiple speakers, along with background noise, environmental interference, and hardware imperfections.
Since the number of speakers, denoted by $N$, is unknown, the signal can be modeled as:
$x(t) = \sum_{i=1}^N s_i(t) + n(t)$, 
where $s_i(t)$ represents the voice signal from voice source $i$ and $n(t)$ represents the noise. 
The objective of this model is to estimate the number of voice sources and separate them.
i.e.,
$[\hat{s}_1(t), \hat{s}_2(t), \ldots, \hat{s}_{\hat{N}}(t)] = f_\theta(x(t))$, 
where $f_\theta(\cdot)$ represents this model parameterized by $\theta$, $\hat{N}$ is the number of estimated voice sources, and
$\hat{s}_i(t)$ is the estimated voice signal from source $i$.

\textbf{Why Self-Supervised Learning?}
Mixed voice signals can be separated through self-supervised learning by leveraging the inherent statistical regularities in audio mixtures.
Human speech tends to be sparse in the time-frequency domain, meaning different speakers often dominate different spectrogram regions. This sparsity allows a model to infer which components belong together based on co-occurrence and consistency across mixtures.
%
In self-supervised training, models can learn from ``mixtures of mixtures'' without needing ground-truth sources. The model separates each input into components and uses a reconstruction loss to check whether these components can be recombined to recover the original mixtures. This remixing constraint acts as a form of supervision, guiding the model to discover meaningful representations of individual sources.
Over time, the model learns to identify and disentangle recurring speech patterns, enabling effective source separation purely from the structure of the input data.

%

\textbf{Main Idea:}
Our design was inspired by the Mixture Invariant Training (MixIT) \cite{wisdom2020unsupervised}, which enables speaker separation using only the demodulated audio stream.
Training samples are constructed by adding two segments of the audio to form a ``mixture of mixtures'' (MoM).
A neural network is trained to separate the MoM into a set of latent sources, which are then remixed to approximate the original input mixtures.
The model is optimized using a signal-level loss that promotes source separation without requiring any ground-truth voice data.
Importantly, this method supports a variable number of audio sources, making it well-suited for our application.

\begin{figure}
\centering
 \includegraphics[width=\linewidth]{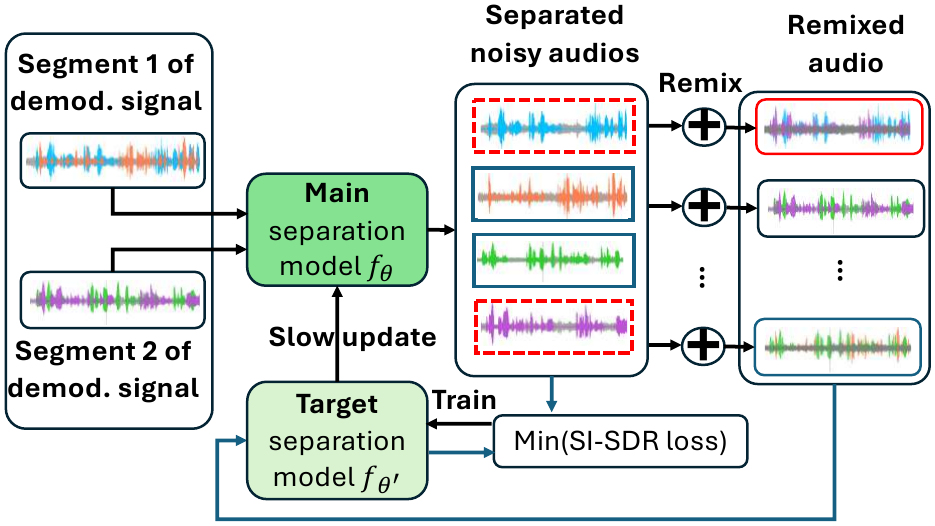}
\vspace{-0.2in}
\caption{Self-supervised training of audio separation model.}
\vspace{-0.15in}
\label{fig:separation_model}
\end{figure}

\textbf{Our Design:}
We use Conv-TasNet \cite{luo2019conv} as the backbone network for both the main and target models, as it has been widely adopted for audio processing tasks.
Fig.~\ref{fig:separation_model} illustrates the self-supervised training process.
Two segments of the demodulated audio stream are selected and separately fed into the main model, which produces two sets of pseudo-source signal estimates.
We then randomly select one signal from each set and apply random weights to generate a remixed signal.
The remixed signal is then fed into the target model.
We compute the scale-invariant signal-to-distortion ratio (SI-SDR) loss between the selected pseudo-source signals generated by the main model and the outputs of the target model, and use this loss to update the target model.
The main and target models share the same architecture, and the parameters of the target model are slowly copied to the main model over time.

Mathematically, denote $x_1(t)$ and $x_2(t)$ as two demodulated signal segments as shown in Fig.~\ref{fig:separation_model}.
Let $\hat{\mathcal{S}}^{(1)} = f_{\boldsymbol{\theta}}(x_1)$ and $\hat{\mathcal{S}}^{(2)} = f_{\boldsymbol{\theta}}(x_2)$ represent the pseudo-source signal sets generated by the main model.
We randomly select one signal from each set, denoted as $\hat{s}_a \in \hat{\mathcal{S}}^{(1)}$ and $\hat{s}_b \in \hat{\mathcal{S}}^{(2)}$, and construct a remixed signal $\tilde{x}(t) = \hat{s}_a(t) + \hat{s}_b(t)$.
The target model then produces an estimated source set $\hat{\mathcal{S}}' = f_{\boldsymbol{\theta'}}(\tilde{x})$.
The loss function is defined as:
\begin{equation}
\mathcal{L}_{\mathrm{sep}} = \min_{\pi \in \mathcal{P}} \sum_{j} \ell \left( \hat{s}'_j(t), \hat{s}_{\pi(j)}(t) \right),
\end{equation}

where $\ell(\cdot, \cdot)$ denotes the scale-invariant signal-to-distortion ratio (SI-SDR) loss, and $\mathcal{P}$ is the set of all permutations over the selected pseudo-source signals.
This loss trains the target network with parameters $\boldsymbol{\theta'}$, which periodically update the main network via
$\boldsymbol{\theta} \leftarrow \lambda \cdot \boldsymbol{\theta} + (1 - \lambda) \cdot \boldsymbol{\theta'}$,
where $\lambda$ is an empirical updating rate.

For this design, we have the following remarks.
\textit{First}, we use a main and target model pair to improve training stability. Direct updates to the main model led to unstable convergence in our experiments. This issue is well known in reinforcement learning, where a target network is used to mitigate instability.
\textit{Second}, to accelerate training, we pre-train the main model on out-of-domain (OOD) speech datasets LibriMix \cite{cosentino2020librimix}, and initialize the target model randomly. The main model remains frozen and receives periodic EMA updates with high momentum to preserve stability.
\textit{Third}, since the loss is computed over all permutations of separated signals, the computational cost grows rapidly with the number of speakers. We therefore limit the number of speakers to two in our experiments.

\vspace{-0.05in}
\subsection{Self-Supervised Denoising Model}
\vspace{-0.05in}

\begin{figure}
\centering
 \includegraphics[width=\linewidth]{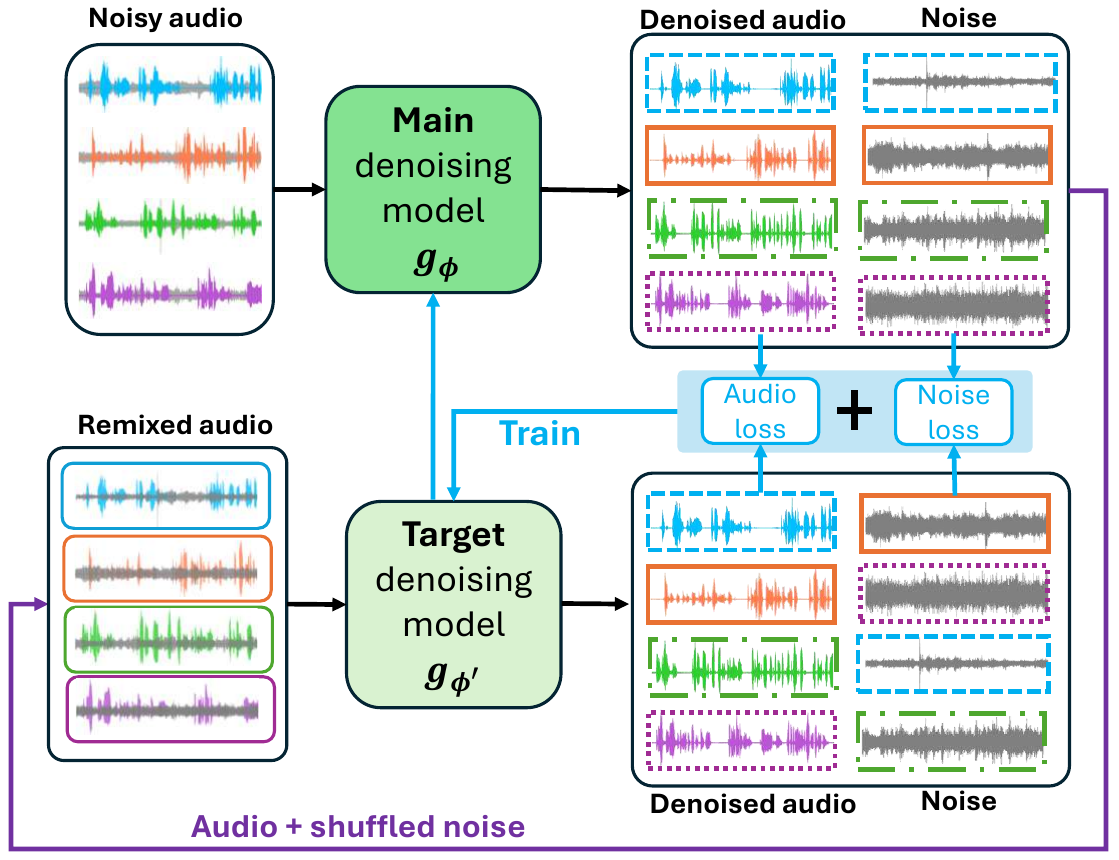}
 \vspace{-0.15in}
\caption{Self-supervised training of audio denoising model.}
\vspace{-0.15in}
\label{fig:denoising_model}
\end{figure}

\textbf{Main Idea:}
Noise and voice signals exhibit distinct characteristics in the temporal domain.
Speech typically follows structured, periodic patterns corresponding to phonemes and syllables, while noise is often unstructured, irregular, or statistically random. These differences enable a model to learn discriminative features that separate speech from noise. By training on diverse mixtures, the model can recognize and exploit these temporal patterns to effectively isolate voice signals from noise.
Our design drew inspiration from the self-training speech enhancement approach proposed in \cite{tzinis2022remixit}.
The core idea is to remix the signal estimates with permuted noise estimates to generate pseudo-labels, which are then used for the self-supervised training of the denoising model.

\textbf{Our Design:}
Fig.~\ref{fig:denoising_model} shows the self-supervised training process for our denoising model.
Similar to the separation task, we adopt a main-target model pair for signal denoising.
The main and target models share the same backbone network architecture, e.g., SuDoRM-RF \cite{tzinis2020sudo} in our experiments.
The main model is used to separate a batch of noisy signals, producing corresponding signal and noise estimates.
Next, we \textit{shuffle} the noise estimates and remix them with the signal estimates to generate new inputs for the target model.
The target model then produces new signal and noise estimates.
We compute the loss by comparing the signal and noise estimates from the main and target models and use this loss to train the target model.
The parameters of the target model are then used to slowly update the main model parameters.
To accelerate the training process, we pre-train the main model using the OOD speech dataset DNS \cite{reddy2020interspeech}, while initializing the target model with randomized parameters. To prevent degradation from noisy updates, we adopt a high-momentum EMA to slowly refine the main model parameters over time.

Mathematically, denote $\hat{s}_i(t)$ as the estimated signal from a single source (i.e., output of the separation model, where $i = 1, 2, \ldots, N$. 
Let $\hat{s}_i(t) = \hat{c}_i(t) + \hat{n}_i(t)$, where $\hat{c}_i(t)$ and $\hat{n}_i(t)$ are the signal and noise estimates generated by the main model. 
Denote $p(\cdot)$ as a permutation function for the integers in $\{1, 2, \dots, N\}$.
Then, we shuffle the noise estimates and mix them with the signal estimates:
$\tilde{s}_i(t) = \hat{c}_i(t) + \hat{n}_{p(i)}(t)$, 
which is fed to the target model. 
Denote $\tilde{c}_i(t)$ and $\tilde{n}_i(t)$ as the signal and noise estimates from the target model.
Then, we calculate the loss function via:
\begin{equation}
\mathcal{L}_{\mathrm{den}} = \sum_{i=1}^N 
\bigg[\ell \big( 
\hat{c}_i(t), \tilde{c}_i(t) \big)
+ 
\ell \big( 
\hat{n}_{p(i)}(t), \tilde{n}_i(t) \big)
\bigg].
\end{equation}
This loss function is used to update the target model. 
Periodically, we use the target network parameters to update the main network parameters by:
$\boldsymbol{\phi} \leftarrow \lambda' \cdot \boldsymbol{\phi} + (1 - \lambda') \cdot \boldsymbol{\phi'}$,
where $\lambda'$ is an empirical updating rate.

\textbf{Feedback of Denoised Signal:}
As shown in Fig.~\ref{fig:system_arc}, the denoised signal can be fed back to the separation model to generate improved signal mixtures for continual training. 
After the denoising model estimates enhanced speech streams from noisy inputs, these outputs are used to construct new synthetic mixtures for the separation model. This feedback mechanism enables the separation model to train on cleaner and more structured input signals, improving its ability to identify and disentangle overlapping sources.

This design is grounded in the concept of distributional alignment \cite{xie2020self}. 
Noisy pseudo-labels from early separation outputs introduce inconsistencies that may hinder learning. By applying denoising prior to remixing, the feedback loop provides higher-quality training targets that better reflect the underlying source structure. The resulting mixtures exhibit clearer speech patterns with increased sparsity in the time-frequency domain, which promotes convergence and facilitates more accurate source decomposition.

Moreover, this feedback encourages mutual co-adaptation between the denoising and separation models. As the denoiser improves the quality of its output, the separator receives progressively cleaner mixtures, allowing both models to refine their parameters in a collaborative and self-reinforcing manner.

%% file: 5_evaluation.tex
To assess the performance of \pname, we conduct a comprehensive evaluation that spans both hardware and software components. Our evaluation seeks to answer the following key questions:

\begin{itemize}
\item \textbf{Effectiveness:} Can \pname reliably recover intelligible speech from raw RF-demodulated signals in multi-speaker acoustic environments?
\item \textbf{Robustness:} How does \pname perform under varying deployment conditions, including different distances, orientations, and environmental noise?
\item \textbf{Adaptability:} Does the system maintain performance when the tag is deployed at diverse locations or when speakers move dynamically within the space?
\end{itemize}

\vspace{-0.05in}
\subsection{Implementation}
\vspace{-0.05in}
\begin{figure}
\centering
 \includegraphics[width=0.875\linewidth]{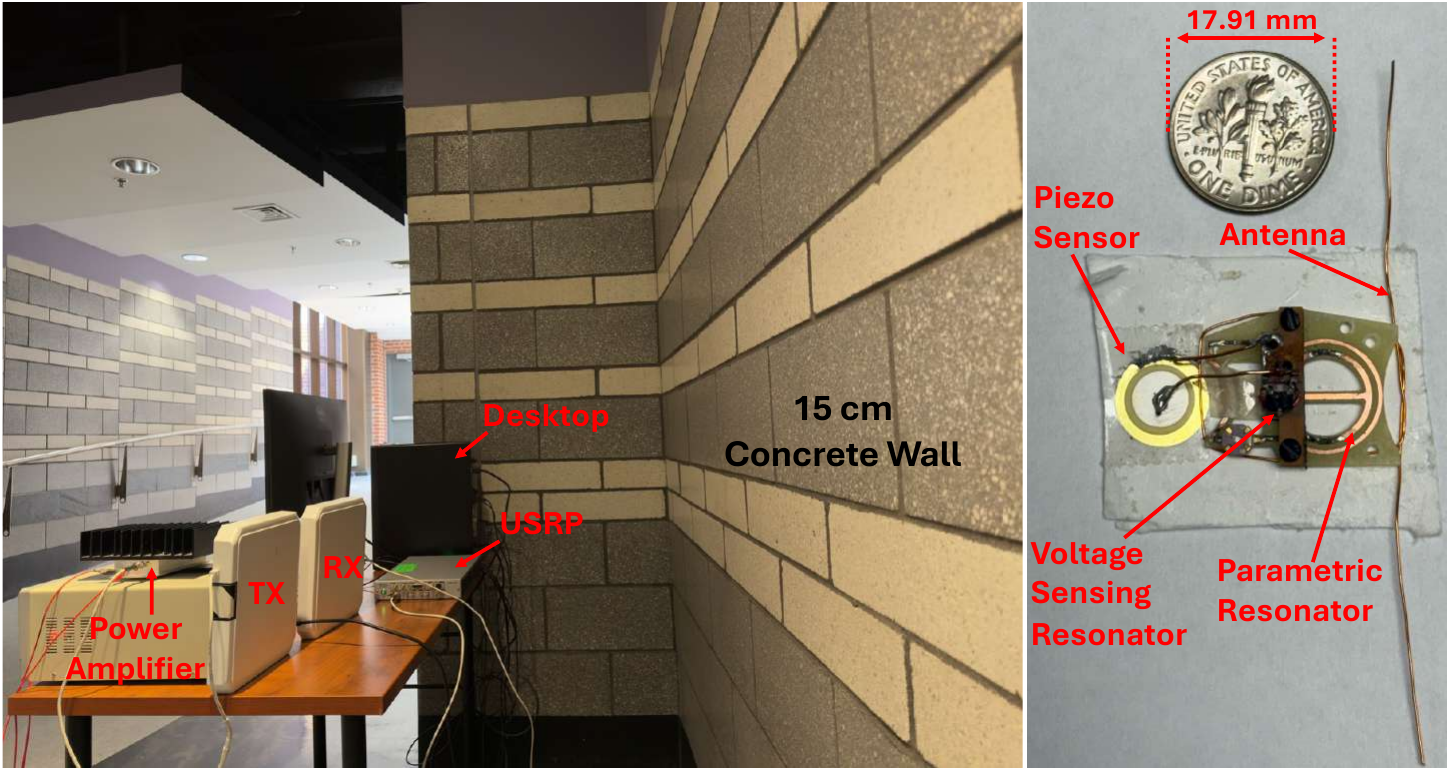}
\vspace{-0.05in}
\caption{RF reader and backscatter tag.}
\vspace{-0.25in}
\label{fig:exp_setup}
\end{figure}

\textbf{RF Tag.}
Fig.~\ref{fig:exp_setup} shows our custom-designed RF tag comprising four components: a Voltage Sensing Resonator (VSR), a Parametric Resonator (PR), a piezoelectric sensor, and a half-wave dipole antenna. 
The VSR uses an $\infty$-shaped coil (32-G copper wire, 5 turns and 1 turn on two rods spaced 1.8\,mm apart), terminated by a bipolar junction transistor (BJT) whose base and emitter connect to the piezoelectric sensor. Operating in the active region, the BJT serves as a voltage-variable capacitor, enabling frequency modulation of the VSR via piezo-induced voltage.  
The PR is a circular planar resonator on a 0.8-mm G10 substrate (13.5/14.5\,mm inner/outer diameters), incorporating varactor diodes (9.1\,pF) across split gaps to support dual-mode resonance (circular and butterfly modes).  
A half-wave dipole antenna wraps around the PR’s perimeter for loose inductive coupling, boosting energy transfer without a direct electrical connection.  
Together, the PR acts as a frequency-selective amplifier that upconverts the VSR’s modulated signal into a detectable backscatter waveform.

\textbf{RF Reader.}
The RF reader setup, as shown in Fig.~\ref{fig:exp_setup}, consists of a USRP N310, a power amplifier (PA), two directional antennas, and a host PC.
The reader continuously transmits an excitation signal at 915\,MHz with a power output of 30\,dBm.
In response, the tag generates a frequency-modulated backscatter signal centered at 515\,MHz, corresponding to the circular-mode resonance frequency of the PR.

\textbf{Separation Model.}
We adopt Conv-TasNet \cite{luo2019conv} as the backbone network for source separation.
The main model is pre-trained on LibriMix \cite{cosentino2020librimix} (5s segments), using STFT (512-point window, 128 hop, Hann window) with mixture-consistent masks. 
After in-domain data collection, we employ a main-target model setup: the main model generates pseudo-labels via cross-mixture remixing, and the target model is trained with permutation-invariant negative SNR loss (max 25\,dB). 
Training uses Adam (lr = 0.001, $\beta_1 = 0.9$, $\beta_2 = 0.999$), batch size 128, gradient clipping, and early stopping.

\textbf{Denoising Model.}
We adopt SuDoRM-RF~\cite{tzinis2020sudo} as the backbone network for denoising.
The main model is pre-trained on DNS~\cite{reddy2020interspeech} to estimate speech/noise components. 
These are shuffled and remixed to train a randomly initialized target model with negative SI-SDR loss. 
We apply EMA updates from the target to the main model after each step for stability. Training uses 16\,kHz audio, Adam optimizer (initial lr = $10^{-3}$, halved every 6 epochs), batch size 2, and standardized inputs.

\subsection{Experimental Setup and Evaluation Metrics}
As illustrated in Fig.~\ref{fig:exp_setup}, we position the RF reader outside the target room, separated by a 15~cm thick concrete wall that is impermeable to mmWave signals. 
The RF tag is discreetly deployed inside the room and placed in inconspicuous locations to avoid detection.
To evaluate the quality of recovered voice signals, we employ a combination of signal-level and perceptual-level metrics.
Signal-level metrics offer objective measures of reconstruction fidelity, while perceptual metrics reflect intelligibility and quality as perceived by human listeners.
Specifically, we use the following metrics for evaluation.

\begin{itemize}[leftmargin=0.15in]
\item
\textbf{SI-SDR (Signal-to-Distortion Ratio):} A signal-level metric that quantifies the fidelity of the reconstructed signal relative to the distortion, invariant to global scaling.

\item
\textbf{LLR (Log-Likelihood Ratio):} A signal-level metric that measures spectral distortion between the clean and enhanced signals using linear predictive coding coefficients.

\item
\textbf{STOI (Short-Time Objective Intelligibility):} A perceptual-level metric ranging from 0 to 1, which assesses speech intelligibility by comparing the short-time spectral envelopes of the clean and processed signals.

\item
\textbf{PESQ (Perceptual Evaluation of Speech Quality):} A perceptual-level metric ranging from 1 to 4.5, designed to quantify the overall perceptual quality of speech based on comparisons with a reference signal.
\end{itemize}

For LLR, lower values indicate better performance.
For SI-SDR, STOI, and PESQ, higher values correspond to improved recovery quality. 

\vspace{-0.05in}
\subsection{Ablation Study}
\vspace{-0.05in}

\input{tables/ablation}
To quantify the contribution of each software component in \pname, we conduct a comprehensive ablation study focusing on three key design elements: (i) pre-training, (ii) exponential moving average (EMA) updates, and (iii) the feedback loop between the separation and denoising modules.
All experiments are performed under the same dataset and conditions used in the main evaluation.

As shown in Table~\ref{tab:ablation}, pre-training provides a strong initialization that significantly boosts performance across all metrics.
In the absence of pre-training, we observe a substantial drop in SI-SDR (from 10.87\,dB to 7.56\,dB), along with corresponding declines in intelligibility and perceptual quality.
In addition, the feedback loop, which enables continual co-adaptation between the denoiser and separator, further enhances the overall system performance.

\vspace{-0.05in}
\subsection{Impact of Victim-to-Tag Distance}
\vspace{-0.05in}

\input{tables/eva_distance}
To assess how speaker-to-tag distance impacts \pname's ability to recover intelligible speech, we systematically vary the distance from 50\,cm to 600\,cm in 50\,cm increments, while keeping the RF reader and tag positions fixed. 
The tag is concealed beneath a table to emulate a realistic covert deployment.
As shown in Table~\ref{tab:tag_dis}, \pname maintains strong performance when the speaker is within 2.5\,m of the tag, achieving median SI-SDR above 7.06\,dB, LLR below 0.51, STOI above 0.67, and PESQ exceeding 2.83.
Under these conditions, the perceptual quality of the recovered speech remains nearly indistinguishable from the original.
As the distance increases beyond 2.5\,m, all metrics exhibit a gradual decline, reflecting diminished acoustic excitation at the tag and reduced signal fidelity.


\begin{figure}
\centering
 \includegraphics[height=1.35in]{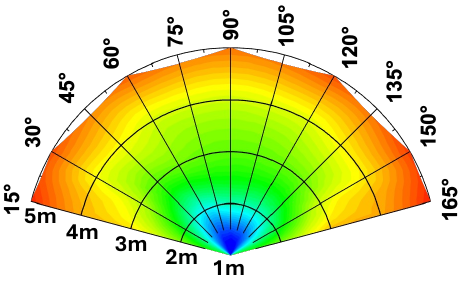}
 \hspace*{0.1in}
 \includegraphics[height=1.35in]{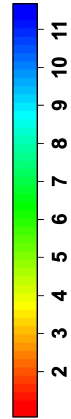}
\caption{Measured SI-SDR Across \pname Orientations.}
\vspace{-0.15in}
\label{fig:orien}
\end{figure}

\vspace{-0.05in}
\subsection{Impact of Reader-to-Tag Orientation}
\vspace{-0.05in}

A key design element of \pname is its circular parametric resonator, which efficiently harvests RF energy across a wide range of incident angles.
While energy transfer is theoretically maximized when the reader is perpendicular to the resonator, indoor multipath reflections help compensate for angular misalignment.

To evaluate orientation sensitivity, we vary the angle between the RF reader and tag from 15\textdegree\ to 165\textdegree\ in 15\textdegree\ steps, across multiple reader-to-tag distances (1–5~m).
Recovered audio is evaluated using the SI-SDR metric.

As shown in Fig.~\ref{fig:orien}, \pname consistently recovers speech at all tested angles.
Although SI-SDR slightly varies with angles, the degradation is far less than that caused by increased distance.
These results show that the resonator's circular geometry and indoor reflections enable reliable energy harvesting and voice recovery without precise reader alignment.

\begin{figure}
\centering
 \includegraphics[width=1\linewidth]{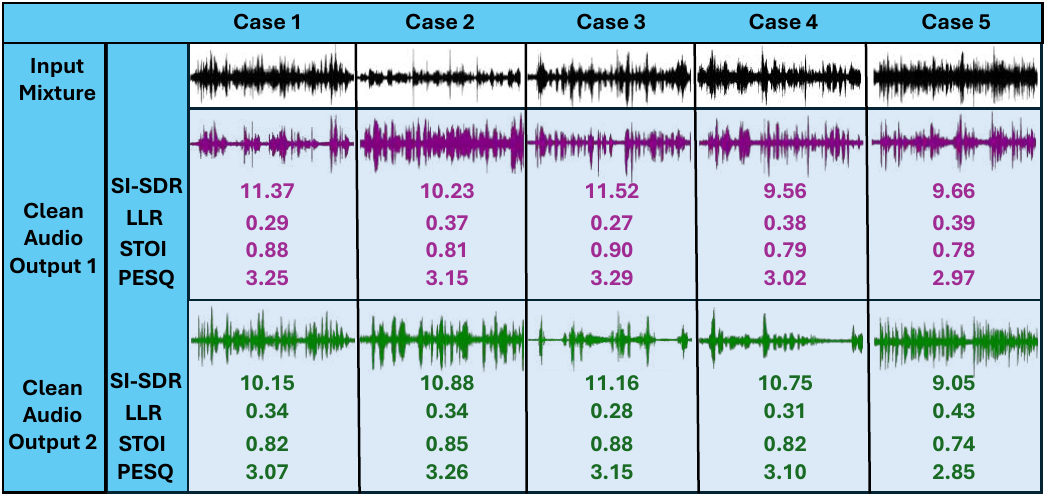}\vspace{-0.05in}
\caption{Impact of noise on separation and denoising.}
\vspace{-0.25in}
\label{fig:env_noise}
\end{figure}

\vspace{-0.05in}
\subsection{Impact of Environmental Noise}
\vspace{-0.05in}
To assess \pname’s resilience in noisy environments, we introduce controlled background disturbances during voice capture and evaluate their impact on audio recovery quality. 
Specifically, we simulate five realistic acoustic scenarios in which two participants speak simultaneously while different environmental noise sources are active.
The noise types include: 
(i) background music played through a speaker,
(ii) keystrokes from a mechanical keyboard,
(iii) periodic smartphone notification alerts,
(iv) prerecorded urban street noise, and 
(v) speech from online videos played on a mobile phone.

As illustrated in Fig.~\ref{fig:env_noise}, each case corresponds to one of the above noise conditions.
\pname maintains strong performance under background music and smartphone alerts, with minimal degradation in speech separation.
In the keystroke scenario, the system effectively removes the pulse-like tapping sounds while preserving speech intelligibility.
Performance declines moderately in the presence of urban noise and online video playback, primarily due to the introduction of competing speech-like signals that overlap with target voices.
Nevertheless, even in these more challenging settings, \pname continues to recover intelligible speech, demonstrating strong resilience to diverse acoustic interference.

\vspace{-0.05in}
\subsection{Impact of Tag Placement}
\begin{figure}
\centering
 \includegraphics[width=1\linewidth]{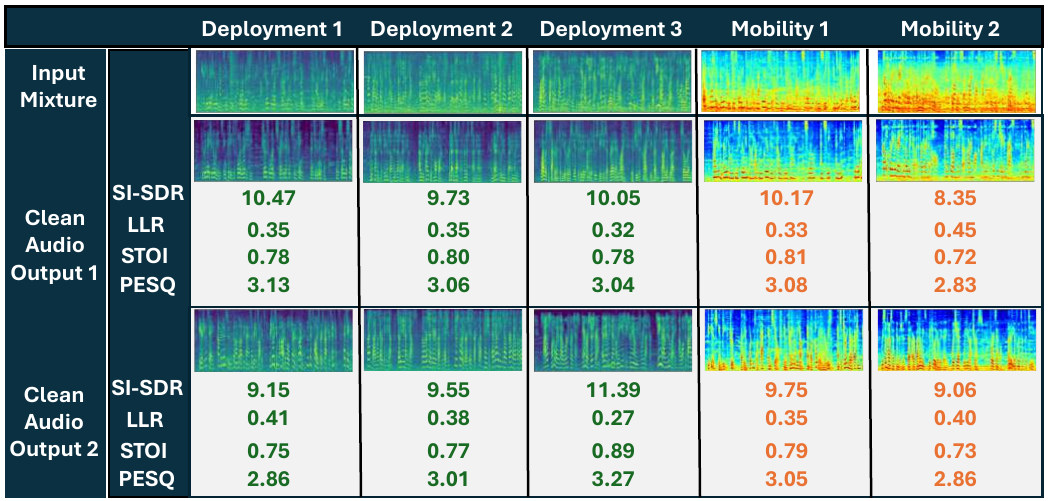}
\caption{Mel Spec. across tag locations and speaker mobility.}
\vspace{-0.2in}
\label{fig:loc_mob}
\end{figure}

\vspace{-0.05in}
To examine how tag placement affects \pname’s performance, we test three typical concealment scenarios: 
(i) beneath a desk, 
(ii) inside an inconspicuous box placed on a shelf, and
(iii) mounted on the backside of a furniture block.
As illustrated in Fig.~\ref{fig:loc_mob}, Deployment1 through 3 correspond to the above cases, respectively. 
Across all settings, \pname consistently achieves intelligible speech recovery, demonstrating strong robustness to placement variation.

Among the three, enclosing the tag in a box slightly reduces performance, likely due to attenuation of acoustic and RF signals.
Mounting the tag on the backside of a furniture block performs better than placing it under a desk, as furniture blocks are more stationary and less affected by human interaction, while tables may experience frequent movement or vibration, introducing interference that degrades signal quality.

\vspace{-0.05in}
\subsection{Impact of Target Movement}
\vspace{-0.05in}
To evaluate the robustness of \pname under target mobility, we design two experimental scenarios involving speaker movement during voice capture:
\textit{(i) Single moving participant.} Two participants remain stationary and speak simultaneously, while a third individual walks continuously within the room, occasionally passing near the tag.
\textit{(ii) Two moving speakers.} Both speaking participants walk around the room while conversing. Their movement paths include varying distances, angles, and intermittent proximity to the tag.

As illustrated in Fig.~\ref{fig:loc_mob}, Mobility1 and Mobility2 correspond to the two scenarios described above.
In Mobility1, where only a non-speaking participant moves, \pname maintains strong performance with no noticeable degradation.
In contrast, Mobility2 introduces significant degradation, likely due to increased variation in speaker position and orientation, which leads to greater signal overlap and spectral ambiguity.
Nevertheless, \pname recovers intelligible speech, demonstrating resilience under complex real-world mobility.

%% file: tables/ablation.tex
\begin{table}
\scriptsize
\centering
\caption{Ablation study.}
\vspace{-0.05in}
\label{tab:ablation}
\begin{tabular}{l|cccc}
\hline
Configuration               & \multicolumn{4}{l}{\ \ \ \ \ \ \ \ \ \ \ Choice} \\ \hline
w/o PT          & \ding{52}     & \ding{56}    & \ding{56}   & \ding{56}   \\
w/ PT           & \ding{56}     & \ding{52}    & \ding{52}   & \ding{52}   \\
EMA Update      & \ding{52}     & \ding{56}    & \ding{52}   & \ding{52}   \\
Feedback Loop   & \ding{52}     & \ding{56}    & \ding{56}   & \ding{52}   \\ \hline
SI-SDR  &   7.56   &   8.75   &   9.42   &   10.87   \\
LLR     &   0.45   &   0.37   &   0.34   &   0.29   \\
STOI    &   0.69   &   0.78   &   0.81   &   0.85  \\
PESQ    &   2.89   &   3.04   &   3.09   &   3.27  
\\\hline
\end{tabular} \vspace{-0.1in}
\end{table}

%% file: tables/eva_distance.tex
\begin{table}
\centering
\scriptsize
\caption{\pname's performance at different distances.}
\vspace{-0.05in}
\label{tab:tag_dis}
\begin{tabular}{c|c|c|c|c}
\hline
Distance & SI-SDR$^\uparrow$ & LLR$^\downarrow$ & STOI$^\uparrow$ & PESQ$^\uparrow$ \\ \hline

\!\!\!\!50 cm\!\!\!\!             & 13.75          & 0.23       & 0.92        & 3.46         \\ \hline
\!\!\!\!100 cm\!\!\!\!            & 12.17          & 0.28       & 0.89        & 3.32           \\ \hline
\!\!\!\!150 cm\!\!\!\!            & 10.24          & 0.31       & 0.84        & 3.19          \\ \hline
\!\!\!\!200 cm\!\!\!\!            & 7.93          & 0.47        & 0.71        & 2.92                 \\ \hline
\!\!\!\!250 cm\!\!\!\!            & 7.06          & 0.51        & 0.67        & 2.83             \\ \hline
\!\!\!\!300 cm\!\!\!\!            & 5.22          & 0.69        & 0.56        & 2.51               \\ \hline
\!\!\!\!350 cm\!\!\!\!            & 4.49          & 0.77        & 0.53        & 2.35               \\ \hline
\!\!\!\!400 cm\!\!\!\!            & 4.15          & 0.85        & 0.50        & 2.31           \\ \hline
\!\!\!\!450 cm\!\!\!\!            & 2.77          & 0.96        & 0.39        & 1.76               \\ \hline
\!\!\!\!500 cm\!\!\!\!            & 2.13          & 1.15        & 0.34         & 1.62             \\ \hline
\!\!\!\!550 cm\!\!\!\!            & 1.84         & 1.37        & 0.31         & 1.48         \\ \hline
\!\!\!\!600 cm\!\!\!\!            & 1.13          & 1.44        & 0.25        & 1.17           \\ \hline
\end{tabular}\vspace{-0.1in}
\end{table}

%% file: 6_related_work.tex
\vspace{-0.05in}
\subsection{Acoustic eavesdropping}
\vspace{-0.05in}
Recent advances have shown that speech can be recovered without traditional microphones, using various wireless and sensing modalities. These systems differ in sensing targets, signal processing, and hardware platforms.


A large body of work employs millimeter-wave (mmWave) radar to detect voice-induced vibrations from human tissue or nearby surfaces. For example, Wavesdropper \cite{wang2022wavesdropper} and Radio2Text \cite{zhao2023radio2text} utilize commercial mmWave sensors to extract vibrations from the throat or environmental surfaces. AmbiEar \cite{zhang2022ambiear} and RADIOMIC \cite{ozturk2022radiomic} extend this to passive objects, while mmEve \cite{wang2022mmeve} targets earpiece emissions. Shi et al. \cite{shi2023privacy} introduce a phased MIMO array that improves directionality and enables limited through-wall sensing, though it still relies on precise object localization and calibration.

Backscatter-based systems like RFSpy \cite{chen2024rfspy} and RF-Parrot \cite{yang2024rf} utilize passive tags or conductive paths to capture vibrations, yet depend on resonant structures or wired access. UWB systems such as VibSpeech \cite{wang2024vibspeech} offer multipath resilience but are constrained by transmission power and limited range.

Some systems adopt unconventional sensing mechanisms. Sound of Interference \cite{onishi2025sound} exploits electromagnetic leakage from digital microphones via near-field probes. Meanwhile, mmEve \cite{wang2022mmeve} enhances mmWave sensing via generative denoising and IQ compensation to mitigate motion artifacts.
 
Despite their diversity, existing approaches face three core limitations: (i) mmWave and UWB signals are heavily attenuated by walls; (ii) many systems require powered sensors, precise alignment, or object-specific targeting; and (iii) most rely on supervised learning with clean audio pairs.
In contrast, \pname operates in sub-GHz bands for stronger wall penetration, supports passive and batteryless eavesdropping without alignment, and uses a self-supervised learning framework that removes the need for labeled data.


\vspace{-0.05in}
\subsection{Audio Separation and Enhancement.}
\vspace{-0.05in}
Traditional supervised models like Conv-TasNet~\cite{luo2019conv} and SepFormer~\cite{subakan2021attention} achieve strong performance but rely on clean, labeled data. To relax this requirement, self-supervised methods have emerged. MixIT~\cite{wisdom2020unsupervised} trains on mixtures of mixtures but may over-separate. Follow-ups like TS-MixIT~\cite{zhang2021teacher} and MixCycle~\cite{karamatli2022mixcycle} improve training stability. Denoising methods like RemixIT~\cite{tzinis2022remixit} and Self-Remixing~\cite{saijo2023self} iteratively bootstrap pseudo-labels without clean references.
Most prior works address either separation or denoising in isolation. \pname unifies both through a remixing feedback loop, where the denoised outputs are fed back to refine separation, and the separation results in turn guide further enhancement. This self-reinforcing process enhances stability, mitigates over-separation, and enables robust recovery from degraded RF-demodulated signals, without requiring any labeled audio.

%% file: 7_conclusion.tex
In this paper, we presented \pname, a novel batteryless RF backscatter-based eavesdropping system that enables voice recovery from within a soundproof room. Unlike existing backscatter tags, \pname supports \textit{continuous} voice streaming while operating solely on harvested energy. 
Our design enables reliable audio eavesdropping through frequency-domain self-interference mitigation and ultra-low-power analog FM voice encoding on a compact batteryless tag, with an RF reader using self-supervised models for voice denoising and separation.
Extensive experiments demonstrate the system's ability to recover and separate human speech with high fidelity under realistic conditions.
This work introduces a new class of passive, long-range voice eavesdropping attacks and underscores the need for renewed attention to RF-based privacy threats.
